\documentclass[runningheads]{llncs}
\usepackage[T1]{fontenc}
\usepackage{graphicx}
\usepackage{xcolor}
\usepackage{listings}

\definecolor{codegreen}{rgb}{0,0.6,0}
\definecolor{codegray}{rgb}{0.5,0.5,0.5}
\definecolor{codepurple}{rgb}{0.58,0,0.82}
\definecolor{lightgray}{rgb}{0.95,0.95,0.92}

\usepackage{booktabs}
\usepackage{hyperref}
\usepackage{multirow}
\usepackage{comment}
\begin{document}

\setlength{\tabcolsep}{12pt}
\renewcommand{\arraystretch}{1.1}
\title{Evaluating Context Segmentation in Locally Deployable SLMs for Cybersecurity CTF Tasks\thanks{Accepted at the RAISE 2026 Workshop (ESORICS 2026). Non-archival.}}

\titlerunning{Evaluating Context Segmentation in SLMs for Cybersecurity CTF Tasks}
% If the paper title is too long for the running head, you can set
% an abbreviated paper title here
%
\author{Sebastiano Nordio\inst{1}\orcidID{0009-0004-0586-4673} \and
Michele Lotto\inst{2}\orcidID{0009-0001-9133-4297}}
%
%\authorrunning{F. Author et al.}
% First names are abbreviated in the running head.
% If there are more than two authors, 'et al.' is used.
%
\institute{Independent \\ \email{norseb21@outlook.com} \and University of Genoa, Via All'Opera Pia, 16145 Genoa, Italy \\ \email{michele.lotto@edu.unige.it}}

\maketitle              % typeset the header of the contribution
\begin{abstract}
The proliferation of highly capable open-weight Small Language Models (SLMs) democratizes access to advanced cybersecurity capabilities, posing an escalating risk as these models can bypass proprietary API guardrails when deployed locally. However, SLMs deployed as autonomous agents often struggle with long-horizon, exploratory tasks like cybersecurity Capture The Flag (CTF) challenges due to context bloat and cognitive degradation from accumulated tool-call outputs. To understand and mitigate this cybersecurity threat, we introduce \textit{context segmentation}, a two-level agentic framework that divides complex exploitation tasks into manageable, contextually isolated sub-problems. Evaluating on the \texttt{picoCTF} dataset using memory-constrained \texttt{gemma-4} models, we demonstrate that for the E4B model, our strategy acts as an intelligent search, achieving competitive rewards with superior token efficiency compared to brute-force retries, and successfully solving 18.52\% of tasks that standard agentic execution fails to complete. Code is available at \url{https://github.com/9xeb/context-segmentation}.

\end{abstract}
\section{Introduction}

The integration of Large Language Models (LLMs) into the cybersecurity landscape has introduced a profound paradigm shift, yielding both defensive advantages and severe security challenges. On one hand, LLMs can significantly optimize defensive operations, such as automated vulnerability detection, code analysis, and incident response~\cite{tian2025exploringrolelargelanguage}. On the other hand, their dual-use nature raises critical concerns, as these same capabilities can theoretically be weaponized to generate malicious payloads, automate social engineering, and identify vulnerabilities~\cite{gupta2023chatgptthreatgptimpactgenerative}. Despite these theoretical risks, a practical bottleneck has historically existed: foundational studies demonstrated that only massive, state-of-the-art proprietary models possessed the reasoning and long contextual planning required to successfully execute complex, multi-step cybersecurity tasks, such as autonomous web hacking~\cite{fang2024llmagentsautonomouslyhack}. Consequently, the immense computational overhead and API-gated access to these frontier models acted as a natural safeguard. This high barrier made it practically impossible for malicious actors with limited hardware resources to conduct fully automated, large-scale cyberattacks using language models. However, the landscape is rapidly evolving; the continuous proliferation of highly capable open-weight models such as the DeepSeek-R1~\cite{guo2025deepseek}, Gemma4~\cite{gemma4_2026_docs} and Qwen3~\cite{yang2025qwen3} families has democratized access to advanced problem-solving and multi-stage attack analysis~\cite{guo2025deepseek,air2026languagemodelsautonomouslyhack}. Recent demonstrations have proven that models running on consumer hardware can now autonomously discover vulnerabilities, execute exploits, and even self-replicate across network hops without human oversight~\cite{air2026languagemodelsautonomouslyhack}. Because these models can be fine-tuned, uncensored~\cite{arditi2024refusallanguagemodelsmediated}, and deployed locally on consumer-grade hardware, they bypass the safety guardrails of proprietary APIs~\cite{cheng2025weaponization}. This presents a tangible and escalating risk, as the advancing capabilities of these open models could be actively exploited by attackers to automate complex cyber threats without oversight or resource constraints. This automation is made possible by integrating these models into agentic frameworks~\cite{xiesurvey2026}, which transforms LLMs from storytellers to entities capable of autonomously navigating the real world. As a side effect, during the agentic loop, despite LLMs only generating strings, real code is executed under their control by the harness. This means that LLMs paired with tools can now do anything a computer program does, including using email, controlling robotic motors, and cracking IT systems via CLI. 

To understand and mitigate the cybersecurity threat of autonomous, LLM-powered agents, it is essential to evaluate the true offensive potential of locally deployable open-weight Small LMs (SLMs). In this paper, we propose a novel methodology designed to improve the performance of these smaller models in complex cybersecurity scenarios. Our approach introduces \textit{context segmentation}, an automated prompt-level `divide and conquer' strategy that breaks down intricate exploitation tasks into manageable, context-specific sub-problems and processes them sequentially. Specifically, the primary contributions of this work are as follows:

\begin{itemize}
    
\item \textbf{Token-Efficient Agentic Execution:} We demonstrate that while brute-force iteration of baseline agentic loops can yield higher absolute rewards, \textit{context segmentation} provides a significantly more token-efficient alternative.

\item \textbf{Resolution of Complex Edge-Cases:} We show that our segmentation strategy intelligently constrains and guides model generation, enabling the successful resolution of complex, edge-case tasks that even the reiterated baseline fails to address entirely.

\item \textbf{Re-evaluation of SLM Threat Models:} We provide empirical evidence that, when equipped with an optimized prompt-level harness, open-weight SLMs can successfully execute more sophisticated exploitation chains. This lowers the safeguard barrier for malicious actors with limited hardware resources to conduct fully automated cyberattacks using SLM.
\end{itemize}

\section{Background}

Small Language Models (SLM) are models designed to maintain the accuracy and/or adaptability of large language models, while being subject to hardware constraints, such as VRAM and computing power~\cite{van2025survey}. The definitions of `small' and `large' are a function of both context and time. For instance, GPT-2 (1.5B) was considered `large' in 2019 but in 2026 is considered smaller than many `small' language models. In cybersecurity context, while SLMs have demonstrated to nearly reach LLMs performance on tasks such as tactical and technical extraction~\cite{YANG2026}, they still lack behind LLMs in tasks such as Phishing Website Detection~\cite{jcp6020048}. The advent of Agents, which leverage the emerging ability of language models to generate structured text to be parsed by other programs, has been pivotal in cybersecurity~\cite{zhang2025cybenchframeworkevaluatingcybersecurity,abramovich2025enigmainteractivetoolssubstantially}, as well as in other fields~\cite{ferrag2026llmreasoningautonomousai}. SLM-powered agents can now generate JSON or YAML payloads containing function call signatures that external programs process to execute code on behalf of the model. The code execution results are then appended to the context and sent back to the language model, which can subsequently call additional tools or provide a final response. While a conversation is technically occurring, a command interpreter is actually reading and answering on the other side instead of a human, enabling the continuous agentic loop. Since exploitations are typically performed via shell interfaces (e.g., bash), language models are prompted to generate function call signatures containing command chains or scripts to perform exploitations autonomously.

\paragraph{Threat Model: Hardware Constraints.} In evaluating these systems, our threat model specifically assumes an adversary operating under strict hardware limitations. The deployment environment is restricted to standard hardware with limited Video RAM (VRAM) and compute capacities. Consequently, the agent must rely entirely on localized Small Language Models. This constraint dictates that the models cannot leverage massive context windows or trillion-parameter reasoning to solve complex vulnerabilities in a single inference pass, necessitating efficient and tool-assisted interactions through an agentic loop.

\paragraph{Capture The Flag (CTF) Tasks.} To safely and objectively assess the cybersecusity capabilities of these hardware-constrained agents, their function calls must be routed to repeatable sandbox environments. A standard and highly effective metric for this evaluation relies on Capture The Flag (CTF) challenges~\cite{zhang2025cybenchframeworkevaluatingcybersecurity,wang2026cybergymevaluatingaiagents,sanzgomez2025cybersecurityaibenchmarkcaibench,yang2023intercodestandardizingbenchmarkinginteractive}. CTFs are specialized cybersecurity exercises designed to test and develop offensive and defensive security skills. These challenges span various technical domains, including web vulnerability exploitation, reverse engineering, cryptography, and binary exploitation, where the objective is to locate and extract a hidden string of text, known as the `flag', which serves as cryptographic proof that the system was successfully compromised. While originally designed for human cybersecurity learners interacting directly with a terminal, the deterministic environments, isolated nature, and clear success criteria of CTFs make them the ideal benchmarks for assessing an autonomous agent's exploitation capabilities.

While agentic loops enable SLMs to interface with complex terminal environments, they introduce a severe bottleneck in hardware-constrained settings: \textit{context bloat}~\cite{hong2025contextrot}. During multi-step CTF challenges, execution logs, verbose tool outputs, and trial-and-error command outputs rapidly consume token budgets, leading to dead-end execution clutter. Furthermore, SLMs suffer from attention degradation and KV-cache overhead long before reaching their theoretical context limits, leading to instruction drift, hallucinated command syntax, and repetitive failure loops~\cite{zhang2023h2oheavyhitteroracleefficient,xiao2024efficientstreaminglanguagemodels}. \textit{Context segmentation} directly addresses these limitations by automatically applying a `divide and conquer' strategy at the prompt level, breaking down intricate exploitation tasks into manageable, context-specific sub-problems and processing them sequentially.

\section{Methodology}

To translate this conceptual divide-and-conquer strategy into a practical architecture without violating our strict hardware limitations, we must achieve context segmentation using only a single locally hosted SLM. The core intuition is to separate high-level strategic reasoning from low-level tactical execution, ensuring the verbose outputs of terminal commands do not pollute the agent's long-term planning capabilities. 

To achieve this separation using only one underlying model, rather than forcing a single agent to parse its entire history of failed attempts in a single context, our approach divides the cognitive load. A main agent, named \textit{Explorer}, keeps track of summarized, high-level reports on all problem-solving attempts, and uses those reports to generate new attempt instructions. A secondary agent, named \textit{Worker}, is prompted by the main agent and actually performs the single problem-solving attempts as instructed. The \textit{Explorer} receives the problem and is equipped with an agent-as-a-tool named \texttt{run\_strategy}, which accepts a proposed course of action in natural language. Upon invocation, this tool starts the \textit{Worker} as an independent secondary agent loop that executes the proposed strategy from a blank context, while the \textit{Explorer} waits for the tool call to return a result. Unlike the \textit{Explorer}, the \textit{Worker} directly interacts with the execution environment (i.e. a bash CLI tool). Upon termination, the \textit{Worker} generates a summary detailing its actions and its reason for stopping, and the summary is returned to the \textit{Explorer} as a tool output. After each report, the \textit{Explorer} agent evaluates whether to call a new strategy or halt execution. Both the \textit{Explorer} and the \textit{Worker} have access to a \texttt{give\_up} tool, which allows them to signal and justify when they have exhausted all viable options within their respective contexts. Appendix \ref{app:pseudocode} provides the Python pseudocode for our proposed implementation of \textit{context segmentation}.

This approach is supposed to mitigate \textit{context bloat}, by dynamically pruning irrelevant information~\cite{yang-etal-2025-llm-reasoning}. The primary objective of \textit{context segmentation} is to prevent Small Language Model (SLM) agents from becoming disoriented by lengthy sequences of tool interactions, thereby enabling coherent, high-level problem-solving.

\section{Experimental Setup}
\label{sec:experimental_setup}
%It is important to note that the purpose of this research is to study agent harnesses, regardless of models.

\noindent\textbf{Models.}
Our evaluation focuses on agents powered by Instruction-Tuned (it) models, optimized for local deployment. To ensure computational feasibility across varied consumer-grade hardware, all selected models were quantized to 4-bit precision to operate within a strict memory envelope of less than 6 GB of RAM\footnote{Details regarding the utilized hardware and software infrastructure are provided in Appendix \ref{app:hardware-software}.}. Specifically, we evaluate the \texttt{gemma-4-E2B-it}\footnote{\url{https://huggingface.co/unsloth/gemma-4-E2B-it-GGUF}} and \texttt{gemma-4-E4B-it}\footnote{\url{https://huggingface.co/unsloth/gemma-4-E4B-it-GGUF}} variants~\cite{gemma4_2026_docs}, which successfully meet these resource constraints while maintaining competitive reasoning capabilities. See Appendix \ref{app:models} for further discussion on the models utilized.

%Conversely, other candidates within the same parameter class, specifically the \texttt{Qwen-3.5-4B} and \texttt{Qwen-3.5-9B} models, were excluded from final benchmarking. Preliminary testing revealed that these models exhibited a tendency to rapidly expand their internal reasoning process, leading to runaway generation that frequently exhausted the context window, making their use unfeasible on consumer-grade hardware.

\vspace{0.5em}
\noindent\textbf{Task and Environment.}
Because autonomous agents and their associated harnesses represent a relatively nascent area of research, we encountered a scarcity of cybersecurity-oriented benchmarks tailored to evaluate agents rather than standalone models. For instance, Cybench~\cite{zhang2025cybenchframeworkevaluatingcybersecurity} only supports wiring model APIs and precludes custom agent harnesses; CyberGym~\cite{wang2026cybergymevaluatingaiagents} focuses on software vulnerability analysis rather than system exploitation; and CAIBench~\cite{sanzgomez2025cybersecurityaibenchmarkcaibench} is hidden behind a paywall. Overall, the current evaluation ecosystem remains fragmented and model-centric. 

Consequently, we opted to use a modified version of Intercode~\cite{yang2023intercodestandardizingbenchmarkinginteractive} containing functional Capture The Flag (CTF) challenges, which serve as a proxy for real-world interactive problem-solving and tool manipulation. Specifically, we leverage the interactive \texttt{CTFEnv} environment class. The evaluation dataset comprises a diverse set of cybersecurity challenges sourced directly from the \texttt{picoCTF} platform~\cite{183443}. These challenges require the agent to dynamically interact with a bash shell, examine binaries, and retrieve hidden flags. Further details on Intercode are presented in Appendix~\ref{app:intercode}.

\subsection{Strategies}
We evaluate and compare two operational configurations to assess the impact of \textit{context segmentation}.

\vspace{0.5em}
\noindent \textbf{\textit{Plain} (Baseline):} 
This strategy represents a standard agent execution wherein the agent interacts with the environment sequentially during a single execution trial. We implement this as a single \textit{Worker} agent loop, initialized with the problem description as its prompt. A task is marked as successful if the correct flag is submitted, and as failed if the agent `gives up'.

\vspace{0.5em}
\noindent \textbf{\textit{Explorer}:} 
This is our proposed approach. Similar to the baseline, the task is considered successful upon submission of the correct flag and failed if the \textit{Explorer} agent `gives up'. Experimental observations revealed that this strategy underperforms compared to a standard agent loop on very simple problems, as the \textit{Explorer} agent tends to overcomplicate straightforward solutions. To mitigate this, we refined our approach into the \textit{plain+explorer} strategy. Furthermore, we excluded this from our evaluation in favor of the \textit{plain+explorer} strategy. 

\vspace{0.5em}
\noindent \textbf{\textit{Plain+Explorer}:} 
This strategy combines the baseline with our proposed approach. The system initially employs the \textit{plain} strategy; if the single \textit{Worker} agent fails (i.e., gives up), the system falls back to the \textit{Explorer} approach. The success criteria remain consistent: the task succeeds if the correct flag is submitted and fails if the \textit{Explorer} agent concedes.

\vspace{0.5em}
\noindent Further details on strategies are discussed in Appendix~\ref{app:strategies}.

\section{Results}
\label{sec:results}
In \autoref{tab:strategy_comparison}, we compare the mean reward and mean token usage of our \textit{plain+exp-\\lorer} strategy against the \textit{plain} baseline. As shown, our approach achieves a notable improvement in reward, albeit at the cost of significantly higher token consumption. This trade-off occurs because the \textit{plain+explorer} strategy governs the execution of multiple \textit{plain} instances, effectively iterating the baseline approach several times.

\begin{table}[h]
    \caption{Comparison of mean reward and token usage between the \textit{plain} and \textit{plain+explorer} strategies.}
    \label{tab:strategy_comparison}
    \centering
    \begin{tabular}{l r r r r}
    \toprule
    & \multicolumn{2}{c}{\textbf{gemma-4-E2B-it}} & \multicolumn{2}{c}{\textbf{gemma-4-E4B-it}} \\
    \cmidrule(lr){2-3} \cmidrule(lr){4-5}
    \textbf{Strategy} & \textbf{Reward} & \textbf{Tokens} & \textbf{Reward} & \textbf{Tokens} \\
    \midrule
    plain          & 0.14 & 2362.99 & 0.23 & 2694.54 \\
    plain+explorer & 0.19 & 8031.02 & 0.39 & 8652.52 \\
    \bottomrule
    \end{tabular}
\end{table}

Given these initial insights, it is crucial to determine whether the observed gain in reward is merely the result of iterating the baseline approach multiple times. To investigate this, \autoref{tab:best_of_k} compares the mean reward and token usage of our \textit{plain+explorer} strategy against a \textit{plain best-of-$k$} baseline, where the baseline reward is evaluated using a logical OR over the per-task success of $k$ independent iterations (reward@$k$), and the baseline token usage is calculated as the sum of the tokens consumed across all independent iterations. As shown, for the \texttt{gemma-4-E2B-it} model, our approach does not outperform a simple brute-force iteration of the \textit{plain} strategy. Conversely, for the \texttt{gemma-4-E4B-it} model, our \textit{plain+explorer} strategy demonstrates a strong advantage in token efficiency. While it does not exceed the absolute maximum reward achieved by the resource-heavy \textit{plain best-of-5} baseline (0.41), it attains a highly competitive reward of 0.39 using only 8652.52 tokens. For comparison, the \textit{plain+explorer} approach outperforms the \textit{plain best-of-4} configuration in reward (0.39 vs. 0.38) while consuming over 2,200 fewer tokens. This suggests that with a sufficiently capable base model, the \textit{plain+explorer} strategy intelligently guides the generation toward successful outcomes, achieving a highly favorable trade-off between task success and computational cost compared to brute-force retries.

\begin{table}[h]
    \caption{Comparison of reward and token usage. Reward is measured as reward@$k$ for \textit{plain best-of-$k$} configurations, and as the mean reward for \textit{plain+explorer} experiments.}
    \label{tab:best_of_k}
    \centering
    \begin{tabular}{l r r r r}
    \toprule
    & \multicolumn{2}{c}{\textbf{gemma-4-E2B-it}} & \multicolumn{2}{c}{\textbf{gemma-4-E4B-it}} \\
    \cmidrule(lr){2-3} \cmidrule(lr){4-5}
    \textbf{Experiment} & \textbf{Reward} & \textbf{Tokens} & \textbf{Reward} & \textbf{Tokens} \\
    \midrule
    plain best-of-1 & 0.14 & 2420.47 & 0.25 & 2525.10 \\
    plain best-of-2 & 0.20 & 4799.23 & 0.31 & 5262.07 \\
    plain best-of-3 & 0.22 & 7136.43 & 0.35 & 7940.94 \\
    plain best-of-4 & 0.23 & 9400.00 & 0.38 & 10874.63 \\
    plain best-of-5 & 0.25 & 11814.94 & 0.41 & 13472.68 \\
    \midrule
    plain+explorer  & 0.19 & 8031.02 & 0.39 & 8652.52 \\
    \bottomrule
    \end{tabular}
\end{table}

Finally, in \autoref{tab:strategy_outcomes}, we break down the task successes to better understand the overlap and differences between the two strategies. For the more capable \texttt{gemma-4-E4B-it} model, the \textit{plain+explorer} strategy demonstrates a substantial advantage, successfully solving 18.52\% of the tasks that the baseline \textit{plain} strategy fails to answer. In contrast, the baseline uniquely solves only 2.88\% of the tasks, highlighting a strong benefit when employing the explorer configuration.  Conversely, for the \texttt{gemma-4-E2B-it} model, the vast majority of tasks (77.78\%) remain unsolved by either approach. While the \textit{plain+explorer} strategy still recovers more unique tasks than the baseline (7.82\% vs. 3.29\%), the margin is much narrower. This breakdown proves that the exploration mechanism is highly effective at recovering edge cases and guiding generation, but it requires an underlying model with a sufficient baseline capacity to actually solve the problems once they are explored.

\begin{table}[h]
    \caption{Breakdown of task success. The table shows the percentage of tasks solved exclusively by one strategy, by both, or by neither.}
    \label{tab:strategy_outcomes}
    \centering
    \begin{tabular}{l|r r}
    \toprule
    & \multicolumn{2}{c}{\textbf{ Tasks Solved (\%)}} \\
    \cmidrule(lr){2-3}
    \textbf{Outcome} & \textbf{gemma-4-E2B-it} & \textbf{gemma-4-E4B-it} \\
    \midrule
    Solved only by \textit{plain}          & 3.29  & 2.88  \\
    Solved only by \textit{plain+explorer} & 7.82  & 18.52 \\
    Solved by both                         & 11.11 & 20.16 \\
    Solved by neither                      & 77.78 & 58.44 \\
    \bottomrule
    \end{tabular}
\end{table}

\section{Limitations and Future Work}

While our approach successfully enhanced the capabilities of agents powered by locally deployable SLMs in CTF challenges, current results remain preliminary and highlight critical avenues for future research. 

Moving forward, our efforts will revolve around \textbf{prompt optimization} to move beyond our current hand-crafted implementations, refining the \textit{Explorer} and \textit{Worker} system prompts for maximum strategic coherence and execution reliability. This effort will include introducing new roles to capitalize on aggressive context segmentation, ensuring the architecture supports general problem-solving without relying on specific domain assumptions. Alongside these structural improvements, we will implement an \textbf{agentic long-term memory system} to persistently record, retrieve, and reuse successful exploitation strategies. We must also rigorously reassess the implications of the `give-up' tool to determine if it inadvertently inhibits agents and compels premature laziness during repeated failed attempts at exceptionally hard problems. 

Finally, because our 6GB RAM constraint limited us to only two models that reliably completed the current benchmark (as detailed in Appendix~\ref{app:models}), we intend to test a wider variety of models on low-end hardware. This will culminate in comprehensively validating our methodology through \textbf{expanded benchmarking} across more diverse, high-fidelity security domains that extend far beyond CTF challenges and better reflect real-world vulnerability discovery usable with custom harness frameworks.

\section{Conclusion}

In this work, we investigated the capabilities of locally deployable agents powered by open-weight Small Language Models (SLMs) in executing autonomous cybersecurity tasks. Recognizing that standard agentic loops suffer from rapid context bloat and cognitive degradation, we introduced \textit{context segmentation}, a dual-agent framework that dynamically separates high-level strategic reasoning from low-level execution clutter. It successfully resolves complex, long-horizon tasks that single-pass baselines entirely fail to address, achieving a superior trade-off between absolute reward and token efficiency compared to brute-force repetition. This capability effectively lowers the hardware barrier for automated cyberattacks, highlighting a critical and escalating threat in the offensive security landscape. While our empirical results confirm this threat, they also suggest that SLMs possess a latent agentic potential that remains heavily constrained by traditional, manual prompt engineering. The proposed \textit{context segmentation} approach acts as an initial form of automated prompt expansion, demonstrating significant room for optimization.

\begin{comment}
\begin{credits}
\subsubsection{Author Contributions. }\textbf{S.N.:} Conceptualization, Methodology, Software, Writing; \textbf{M.L.:} Data Curation and Analysis, Validation, Supervision, Writing;
\end{credits}
\end{comment}
%
% ---- Bibliography ----
%
% BibTeX users should specify bibliography style 'splncs04'.
% References will then be sorted and formatted in the correct style.
%
\bibliographystyle{splncs04}
\bibliography{mybibliography}

\newpage
\appendix
\section{Pseudo Python code}
\label{app:pseudocode}
The following pseudo-Python code provides a clear view on our proposed implementation:
\begin{lstlisting}
def give_up():
    """
    Final fallback tool. Use this exclusively to terminate
    the search and indicate failure to find alternatives.
    You must call this ONLY as an absolute last resort, 
    strictly after all other available tools, strategies, 
    and reasoning paths have been exhaustively attempted 
    and have conclusively failed.
    """
    return """Give-up acknowledged. Provide a full bullet 
    point recap of what you have done and why you stopped."""

def run_strategy(strategy: str) -> str:
    """
    Run a strategy to solve the problem, 
    using a smart remote agent.

    Args:
        strategy: the strategy to run, in clear, 
        natural language instructions
    
    Returns:
        the outcome of the execution of the strategy
    """
    # A new, clean agent spawns when a strategy is run
    worker_agent = Agent(tools=tools+[give_up])
    response = worker_agent.run(problem=user_prompt+strategy)
    return response
    
def context_segmentation(user_prompt, tools):
    # Explorer run as the main loop, and autonomous calls to
    # run_strategy trigger other sub-agent loops
    explorer_agent = Agent(tools=[run_strategy, give_up])
    return explorer_agent.run(problem=user_prompt)
\end{lstlisting}

\section{Hardware and Software Infrastructure}
\label{app:hardware-software}
All experiments were conducted on consumer-grade hardware. We employed two distinct setups: a primary workstation equipped with an \texttt{AMD Radeon RX 7600 XT (16GB VRAM)} and a secondary node featuring an \texttt{AMD Radeon RX 6600 XT (8GB VRAM)}. On the software side, model inference was performed using the \texttt{llama.cpp} framework~\cite{11247985}, leveraging the \texttt{Vulkan API} (version 1.4.305)\footnote{\url{https://www.vulkan.org/}}. Furthermore, to ensure a secure and reproducible environment, task execution was isolated using \texttt{Docker}\footnote{\url{https://www.docker.com/}} containers, with a dedicated sandbox deployed for each CTF challenge.

\section{Models}
\label{app:models}

Throughout our experiments, we excluded certain candidate models, specifically \texttt{Qwen-3.5-4B}\footnote{\url{https://huggingface.co/unsloth/Qwen3.5-4B}}, \texttt{Qwen-3.5-9B}\footnote{\url{https://huggingface.co/unsloth/Qwen3.5-9B}}, \texttt{Ling-3.0-Tiny}\footnote{\url{https://huggingface.co/bloomer010/Ling-3.0-tiny-GGUF}}, \texttt{G9v3-3B}\footnote{\url{https://huggingface.co/bartowski/ai9stars_G9v3-3B-GGUF}} and \texttt{LFM2.5-8B-A1B}\footnote{\url{https://huggingface.co/unsloth/LFM2.5-8B-A1B-GGUF}} variants~\cite{inclusionAI_ling_3_0_tiny,ai9stars_g9v3_3b,yang2025qwen3,liquidai2025lfm2} from final benchmarking. During preliminary testing, these models, when quantized to 4 bit precision to fit the 6GB RAM envelope, exhibited runaway generation by rapidly expanding their internal reasoning processes, exhausting the context window after doom loops or excessive reasoning, making them unfeasible to run on consumer-grade hardware.

\section{Intercode}
\label{app:intercode}
To evaluate our agents, we employ Intercode's CTFEnv~\cite{yang2023intercodestandardizingbenchmarkinginteractive}, which provides a Dockerized environment containing approximately one hundred picoCTF challenges organized into individual directories. The environment includes a Python SDK to initialize the container and reset challenges, along with two core functions for the agent: executing Bash commands within the container and submitting captured flags.

We integrate these functions as tools within our chosen agent framework, Pydantic AI~\cite{pydantic_ai}, enabling the agent to autonomously navigate and solve challenges within a secure, reproducible environment. Upon submission, the agent's output is evaluated against the ground-truth flag for that specific challenge. If successful, the agent receives positive feedback; otherwise, it may iteratively refine its approach or terminate the attempt.

The included picoCTF tasks range from basic file system exploration to reverse engineering obfuscated code and exploiting network vulnerabilities, with difficulty levels scaling up to those expected of junior cybersecurity professionals. Each task provides the agent with a brief problem description and full access to the isolated challenge directory.

\color{black}
\section{Strategies}
\label{app:strategies}
The original idea of context segmentation came up when trying to develop an engine called agent-compose. When fed a manually written declarative YAML schema containing subproblems, it would solve each subproblem in a separate context and merge the results to solve larger problems. The subproblems could depend on the results of other subproblems. Eventually, despite implementing LLM-driven generation of the YAMLs, performance was still underwhelming. The most successful execution strategy in agent-compose involved a dual agent system that separated high level planning from low level execution, so that part was extracted, the rigid YAML was removed, and it became the focus of this paper.

\section{Additional Results}

In \autoref{tab:raw_results} we present mean reward and mean token usage for individual experiments runs. As discussed in \autoref{sec:results}, each \textit{plain+explorer} run achieves higher reward compared to each \textit{plain} run, albeit at the cost of significantly higher token consumption.
\begin{table}[htbp]
\centering
\caption{Mean reward and Mean token usage for individual experiments runs.}
\label{tab:raw_results}
\begin{tabular}{l|l|rrr}
\toprule
\textbf{Model} & \textbf{Strategy} & \textbf{Run ID} & \textbf{Reward} & \textbf{Tokens} \\
\midrule
\multirow{8}{*}{\textbf{gemma-4-E2B-it}} & \multirow{5}{*}{plain} & 0 & 0.14 & 2420.47 \\
 & & 1 & 0.17 & 2378.77 \\
 & & 2 & 0.12 & 2337.20 \\
 & & 3 & 0.15 & 2263.57 \\
 & & 4 & 0.14 & 2414.94 \\
\cmidrule{2-5}
 & \multirow{3}{*}{plain+explorer} & 0 & 0.16 & 8173.25 \\
 & & 1 & 0.19 & 8247.53 \\
 & & 2 & 0.22 & 7672.27 \\
\midrule
\multirow{8}{*}{\textbf{gemma-4-E4B-it}} & \multirow{5}{*}{plain} & 0 & 0.25 & 2525.10 \\
 & & 1 & 0.23 & 2736.98 \\
 & & 2 & 0.21 & 2678.86 \\
 & & 3 & 0.25 & 2933.69 \\
 & & 4 & 0.21 & 2598.05 \\
\cmidrule{2-5}
 & \multirow{3}{*}{plain+explorer} & 0 & 0.41 & 8814.74 \\
 & & 1 & 0.36 & 8320.30 \\
 & & 2 & 0.40 & 8822.53 \\
\bottomrule
\end{tabular}
\end{table}

\end{document}